\documentclass[traditabstract,letter]{aa}
\pdfoutput=1

\usepackage{graphicx}
\usepackage[version=3]{mhchem}
\usepackage{txfonts}
\usepackage{natbib}
\bibpunct{(}{)}{;}{a}{}{,}
\usepackage{hyperref}
\hypersetup{pdfauthor={M. de Val-Borro et al.},pdftitle={An upper limit
for the water outgassing rate of the main-belt comet 176P/LINEAR observed
with Herschel/HIFI}}
\usepackage[utf8]{inputenc}
\usepackage[T1]{fontenc}
\usepackage{pgfplots}
\pgfplotsset{compat=newest}
\usepgfplotslibrary{external}
\newcommand{\herschel}{{\it Herschel}}
\newcommand\kms{\ifmmode{\rm km\thinspace s^{-1}}\else km\thinspace s$^{-1}$\fi}
\newcommand\ms{\ifmmode{\rm m\thinspace s^{-1}}\else m\thinspace s$^{-1}$\fi}
\newcommand\s{\ifmmode{\rm molec.~s^{-1}}\else mol~s$^{-1}$\fi}
\newcommand{\trans}{$1_{10}$--$1_{01}$}
\newcommand{\rh}{r_\mathrm{h}}
\newcommand{\xne}{x_{n_\mathrm{e}}}
\newcommand{\upper}{4}
\newcommand{\column}{3}

\begin{document}

\title{An upper limit for the water outgassing rate of the main-belt comet
176P/LINEAR observed with \herschel{}/HIFI\thanks{\herschel{} is an ESA
space observatory with science instruments provided by European-led
Principal Investigator consortia and with important participation from
NASA.}}

\author{M.~de~Val-Borro\inst{\ref{inst1}}\fnmsep\thanks{ {\it Current address}:
      Department of Astrophysical Sciences, Princeton University, NJ 08544, USA}
  \and L.~Rezac\inst{\ref{inst1}}
  \and P.~Hartogh\inst{\ref{inst1}}
  \and N.~Biver\inst{\ref{inst2}}
  \and D.~Bockel\'ee-Morvan\inst{\ref{inst2}}
  \and J.~Crovisier\inst{\ref{inst2}}
  \and M.~K\"uppers\inst{\ref{inst3}}
  \and D.~C.~Lis\inst{\ref{inst4}}
  \and S.~Szutowicz\inst{\ref{inst5}}
  \and G.~A.~Blake\inst{\ref{inst4}}
  \and M.~Emprechtinger\inst{\ref{inst4}}
  \and C.~Jarchow\inst{\ref{inst1}}
  \and E.~Jehin\inst{\ref{inst6}}
  \and M.~Kidger\inst{\ref{inst7}}
  \and L.-M.~Lara\inst{\ref{inst8}}
  \and E.~Lellouch\inst{\ref{inst2}}
  \and R.~Moreno\inst{\ref{inst2}}
  \and M.~Rengel\inst{\ref{inst1}}
  }

\titlerunning{An upper limit for the water outgassing rate of 176P/LINEAR}

\institute{
  Max Planck Institute for Solar System Research, Max-Planck-Str.~2, 37191
  Katlenburg-Lindau, Germany\label{inst1}
  \and LESIA, Observatoire de Paris, CNRS, UPMC, Universit\'e
  Paris-Diderot, 5 place Jules Janssen, 92195 Meudon, France\label{inst2}
  \and Rosetta Science Operations Centre, ESAC,
    European Space Agency, 28691 Villanueva de la Ca\~nada,
    Madrid, Spain\label{inst3}
  \and California Institute of Technology, Pasadena, CA 91125, USA\label{inst4}
  \and Space Research Centre, Polish Academy of Sciences, Warsaw,
    Poland\label{inst5}
  \and Institute d'Astrophysique et de Geophysique, Universit\'e de Li\`ege,
    Belgium\label{inst6}
  \and \herschel{} Science Centre, ESAC, European Space Agency, 28691 Villanueva de la Ca\~nada,
    Madrid, Spain\label{inst7}
  \and Instituto de Astrof\'isica de Andaluc\'ia (CSIC), Glorieta de la
  Astronomía s/n, 18008 Granada, Spain\label{inst8}
  }

\date{Received 5 August 2012 / Accepted 25 August 2012}

\abstract{
176P/LINEAR is a member of the new cometary class known as main-belt
comets (MBCs). It displayed cometary activity shortly during its 2005
perihelion passage, which may be driven by the sublimation of subsurface
ices.  We have therefore searched for emission of the \ce{H2O} \trans{}
ground state rotational line at 557 GHz toward 176P/LINEAR with the
Heterodyne Instrument for the Far Infrared (HIFI) onboard the
\herschel{} Space Observatory on UT 8.78 August 2011, about 40 days
after its most recent perihelion passage, when the object was at a
heliocentric distance of 2.58 AU.  No \ce{H2O} line emission was
detected in our observations, from which we derive sensitive 3-$\sigma$
upper limits for the water production rate and column density of
$<~\upper \times 10^{25}$ \s\ and of $< \column \times 10^{10}$
cm$^{-2}$, respectively.  From the peak brightness measured during the
object's active period in 2005, this upper limit is lower than predicted
by the relation between production rates and visual magnitudes observed
for a sample of comets at this heliocentric distance.  Thus, 176P/LINEAR
was most likely less active at the time of our observation than during
its previous perihelion passage.  The retrieved upper limit is lower
than most values derived for the \ce{H2O} production rate from the
spectroscopic search for CN emission in MBCs.}


\maketitle

\section{Introduction}

Classical main-belt asteroids are small bodies that orbit the Sun in low
inclination and low eccentricity orbits between the orbits of Mars and
Jupiter.  Physically, asteroids are thought to be devoid of volatiles,
while comets are icy bodies that become active in the inner solar system
thanks to the sublimation of ices, mostly water.  Comets originate in
the outskirts of the solar system beyond the snow line, where
temperatures in the solar nebula were low enough for water to condense
onto icy grains \citep{1981PThPS..70...35H}.  A new class of bodies has
been discovered recently, the so-called main-belt comets (MBCs), which
have orbital properties that are indistinguishable from standard
asteroids with a Tisserand parameter with respect to Jupiter that is
greater than three, and they display cometary activity in the form of a
dust tail during part of their orbit.  Numerical simulations have shown
that these objects are not comets from the Kuiper Belt or Oort Cloud
that have been recently transferred to orbits within the main belt, but
instead are most likely formed in situ at their current locations
\citep{2002Icar..159..358F}.

Theoretical models suggest that the snow line was initially close to the
Mars orbit due to the absorption of stellar radiation by dust
\citep{2000ApJ...528..995S,2006ApJ...640.1115L}. Therefore, objects
formed at their current locations in the outer asteroid belt may have
been able to accumulate water ice in subsurface reservoirs, despite
the effect of solar radiation.  Determining the composition of
this class of objects can provide important clues to both the thermal
properties that allow water to survive in subsurface layers and the
distribution of volatile materials in the solar nebula to constrain
planet formation mechanisms. Additionally, MBCs may have played an
important role in the delivery of water and other volatiles to the inner
solar system, including the Earth.

The MBC 176P/LINEAR (hereafter 176P) was discovered in 1999 and
originally categorized as asteroid 118401 LINEAR. This object belongs to
the Themis asteroid family. Cometary activity was reported for this
object around perihelion in 2005 \citep{2011AJ....142...29H} by the
Hawaii Trails project
\citep[HTP;][]{2006Sci...312..561H,2009A&A...505.1297H}.  It displayed
a mean photometric excess of $\sim30$\% during a month-long active
period around its perihelion passage, consistent with an approximate
total dust mass-loss of $\sim 7\times10^4$\ kg
\citep{2011AJ....142...29H}.  Although ice sublimation is expected to
trigger MBC activity, gas emission has never been directly detected in
these objects owing to their low activity, which requires very sensitive
observations.  {\it Herschel} proves to be the most sensitive instrument
for directly observing water in a distant comet
\citep[e.g.][]{2010DPS....42.0304B}.  In this paper we present the
\herschel{} observation of the \trans{} fundamental rotational
transition of \ce{H2O} at 557 GHz in 176P.  This observation is intended
to test the prediction that the observed cometary activity of MBCs is
driven by sublimation of water ices and to constrain the production
process.

\section{Observations}

The MBC 176P was observed with the Heterodyne Instrument for the Far
Infrared \citep[HIFI;][]{2010A&A...518L...6D}, one of the three
instruments onboard the ESA \herschel{} Space Observatory
\citep{2010A&A...518L...1P}, within the framework of the \herschel{}
guaranteed-time key program ``Water and related chemistry in the solar
system'' \citep{2009P&SS...57.1596H}.  HIFI provides very
high-resolution spectroscopy that can resolve the line shape and enable
the determination of accurate production rates
\citep[e.g.,][]{2010A&A...518L.150H}.  176P was the best MBC target in
terms of its visibility close to the perihelion passage and anticipated
line strength to be observed by \herschel{} during the mission lifetime.
It passed its perihelion on 30 June 2011 at a distance of 2.57 AU from
the Sun and was observed by \herschel{} on UT 8.78 August 2011 with a
total on-target integration time of 4.8 hours, when it was at a
heliocentric distance of 2.58 AU and a distance of 2.55 AU from the
satellite (\herschel\ ObsID 1342225905).  The object was tracked using
an up-to-date ephemeris provided by the JPL's Horizons system.

The line emission from the fundamental (\trans) rotational transition of
ortho-water at 557 GHz was searched for in the upper sideband of the
HIFI band 1a mixer.  The observation was performed in the
frequency-switching observing mode with a frequency throw of 94.5 MHz,
using both the wide band spectrometer (WBS) and the high-resolution
spectrometer (HRS).  In this observing mode there is no need to observe
a reference position on the sky and the on-target integration time is
maximized.  However, the statistical noise may be underestimated for
observations in frequency-switched mode owing to uncertainties in
baseline removal \citep{2012A&A...544L..15B}.  The spectral resolution
of the WBS is 1 MHz ($\sim0.54$ \kms at the frequency of the observed
line), while the HRS was used in its high-resolution mode with a
resolution of 120 kHz ($\sim0.065$ \kms).  The main beam brightness
temperature scale was computed using a beam efficiency of 0.75  and a
forward efficiency of 0.96.  The folded spectrum was obtained by
averaging the original spectrum with a shifted and inverted copy.
Horizontal and vertical polarizations were averaged, weighted by the
root mean square amplitude, to increase the signal-to-noise ratio.  The
pointing offset of horizontal and vertical polarization spectra is
6\farcs6 in band 1a, corresponding to approximately 20\% of the
half-power beam width at the observed frequency.

\section{Data analysis}

\begin{figure}
  \centering
  \begin{tikzpicture}
    \begin{axis}[xlabel={$v$ [\kms]},
      xmin=-58.02646, xmax=65.67232,
      ymin=0, ymax=1, axis y line=none,
      axis x line*=top,x dir=reverse,width=\hsize, height=4cm]
    \end{axis}
    \begin{axis}[name=original, ylabel={$T_\mathrm{mB}$ [K]}, const
      plot mark right,
      xmin=556.8139,xmax=557.0437696,axis x line=none,
      xticklabel=\empty, width=\hsize, height=4cm]
      \addplot[mark=none] table [x index=0,y index=1]
   	{/home/miguel/HssO/176P/python/average_HRS.txt};
    \end{axis}
    \begin{axis}[at={(original.south)},anchor=north,
      name=imf9, ylabel={$T_\mathrm{mB}$ [mK]},
      enlarge x limits=false,
      scaled y ticks={real:1e-3},
      xticklabel=\empty,ytick scale label
      code/.code={}, width=\hsize, height=3cm]
      \addplot[mark=none] table [x index=0,y index=9]
   	{/home/miguel/HssO/176P/data/HRS-emds.dat};
      \node at (rel axis cs:.1,.8) {IMF 8};
    \end{axis}
    \begin{axis}[at={(imf9.south)},anchor=north, name=imf8,
      ylabel={$T_\mathrm{mB}$ [mK]},
      enlarge x limits=false,
      scaled y ticks={real:1e-3},
      ytick={-8e-3,-4e-3,0,4e-3,8e-3},
      xticklabel=\empty,ytick scale label
      code/.code={}, width=\hsize, height=3cm]
      \addplot[mark=none] table [x index=0,y index=8]
   	{/home/miguel/HssO/176P/data/HRS-emds.dat};
      \node at (rel axis cs:.1,.8) {IMF 7};
    \end{axis}
    \begin{axis}[at={(imf8.south)},anchor=north, name=imf7,
      ylabel={$T_\mathrm{mB}$ [mK]},
      enlarge x limits=false,
      scaled y ticks={real:1e-3},
      xticklabel=\empty,ytick scale label
      code/.code={}, width=\hsize, height=3cm]
      \addplot[mark=none] table [x index=0,y index=7]
   	{/home/miguel/HssO/176P/data/HRS-emds.dat};
      \node at (rel axis cs:.1,.8) {IMF 6};
    \end{axis}
    \begin{axis}[at={(imf7.south)},anchor=north,
      xlabel={$f$ [GHz]}, ylabel={$T_\mathrm{mB}$ [mK]},
      enlarge x limits=false, scaled y ticks={real:1e-3}, ytick
      scale label code/.code={}, width=\hsize, height=3cm]
      \addplot[mark=none] table [x index=0,y index=2]
   	{/home/miguel/HssO/176P/data/HRS-emds.dat};
      \node at (rel axis cs:.1,.8) {IMF 1};
    \end{axis}
  \end{tikzpicture}
  \caption{Original HRS subband 1 spectrum of the \ce{H2O} \trans{} line
  at 556.936 GHz observed on UT 8.78 August (upper panel), and several
  low and high frequency components of the spectrum determined using the
  EMD analysis (four lower panels) with labels indicating the mode
  number.  The vertical axis is the calibrated main beam brightness
  temperature.  The lower horizontal axis is the upper sideband
  frequency, while the upper axis shows the velocity with respect to the
  nucleus's rest frame.  }
  \label{fig:hrs}
\end{figure}
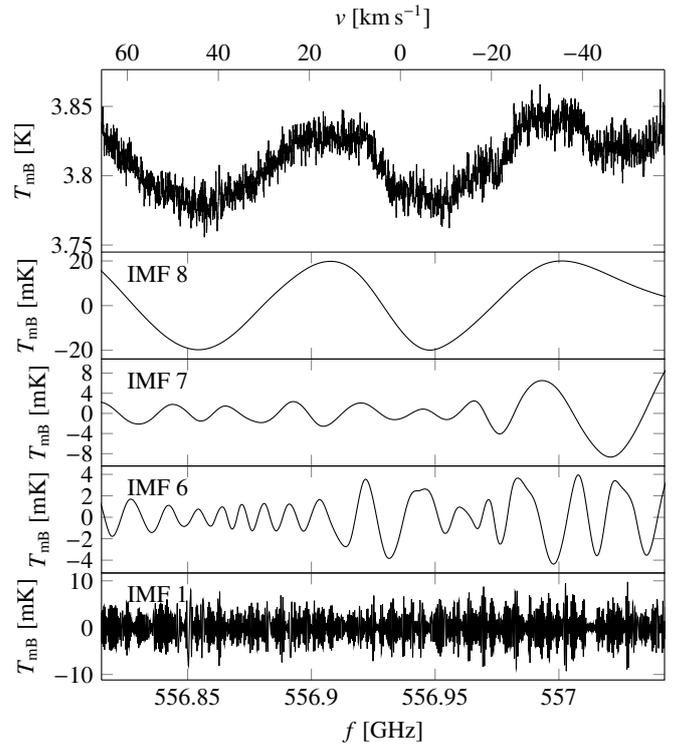

The data analysis was performed using the \herschel{} Interactive
Processing Environment (HIPE) software package
\citep{2010ASPC..434..139O}.  The standard HIFI processing pipeline
v7.1.0 was used to reduce the data to calibrated level-2 data products.
The frequency-switching observing mode introduces a strong baseline
ripple.  To obtain a reliable estimate of the noise present in
the measured data, the  baseline has to be removed, which is usually
accomplished by fitting a linear combination of sine waves  using the
Lomb-Scargle periodogram technique.  Nevertheless, the instrumental
processes responsible for the baseline are in general combinations of
linear distortions of different components in the receiver/spectrometer
subsystems with a small fraction of nonlinear processes, which may cause
an aperiodicity in the ripple.  Analyzing such contaminated signals
by assuming a linear relation among the signal components (a fundamental
assumption for all the Fourier-based techniques) is not always suitable,
depending on the degree of nonlinearity. In this work, we utilize a
relatively novel approach specifically developed for analysis of
aperiodic and nonlinear signals -- the Hilbert-Huang Transform
\citep[HHT;][]{1998RSPSA.454..903E,Huang:1999,2007Geop...72H..29B}. This
approach combines the empirical mode decomposition (EMD) procedure,
which decomposes the original signal into its intrinsic mode functions
(IMFs; representing the different modes of oscillations) with the
Hilbert transform that can be then used in computing the instantaneous
frequencies. The EMD technique extracts all the oscillatory modes,
including all the baseline ripple components (the smooth low-frequency
modes), as well as the highest frequency components, usually the noise.
Another important property of IMFs is that they obey a simple additive
rule to reconstruct the original signal exactly. This makes the EMD
approach an accurate and versatile tool that has been successfully used
in many areas from atmospheric science to cosmology such as denoising and
detrending, as well as the time series analysis tool for identifying
periodic and quasi-periodic features \citep[see][and references
therein]{2004JAtOT..21..599D,2007Geop...72H..29B}.

\begin{figure}
  \centering
  \begin{tikzpicture}
    \begin{axis}[xlabel={$f$ [GHz]},
      xmin=556.921, xmax=556.951,
      xtick={556.93,556.94,556.95},
      ymin=0, ymax=1,
      axis x line*=bottom,
      axis y line=none,
      width=\hsize]
    \end{axis}
    \begin{axis}[xlabel={$v$ [\kms]},
      ylabel={$T_\mathrm{mB}$ [mK]},
      xmin = -8,
      xmax = 8,
      scaled y ticks={real:1e-3},
      ytick scale label code/.code={},
      axis x line*=top,
      x dir=reverse,
      width=\hsize]
      \addplot[mark=none,const plot mark right] table [x index=0,y index=1]
   	{/home/miguel/HssO/176P/python/HRS_HV.dat};
      \addplot[mark=none,color=red,dashed] table [x index=0,y index=1]
	{/home/miguel/HssO/176P/ratran/H2O_q4e25_001.txt};
    \end{axis}
  \end{tikzpicture}
  \caption{High-frequency component of the HRS spectrum of the \ce{H2O}
  \trans{} line at 556.936 GHz observed on UT 8.78 August with overplotted
  synthetic spectrum of the 3-$\sigma$ upper limit shown as the dashed
  line.  The vertical axis is the calibrated main beam brightness
  temperature.  The lower horizontal axis is the upper sideband
  frequency, while the upper axis shows the velocity with respect to the
  nucleus's rest frame.}
  \label{fig:model}
  \end{figure}
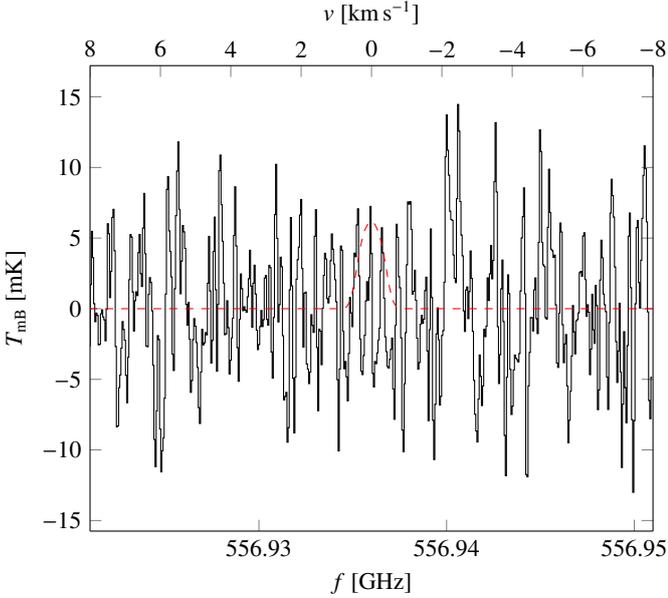

In this work we applied the EMD technique to the measured WBS and HRS
spectra to obtain the highest frequency IMFs (dominated by Gaussian
noise), which are usually the first modes \citep{2004JAtOT..21..599D}.
Some of the modes obtained from the decomposition of the HRS spectrum
are shown in the lower panels of Fig.~\ref{fig:hrs}. Then, the root mean
square (rms) noise of the brightness temperature in each spectrum is
used to derive a 3-$\sigma$ upper limit for the \ce{H2O} production rate
shown in Table~\ref{tbl:q}.  The brightness temperature rms of
the WBS and HRS spectra differ owing to their different spectral
resolutions as predicted by the \herschel\ Observation Planning Tool
(Hspot).  The rms agrees with the value derived using the Lomb-Scargle
periodogram method to determine the frequencies of the baseline.
Typically the Lomb-Scargle implementation requires 20 to 30 sinusoidal
components to achieve a good fit of the baseline ripple, while the HHT
analysis provides a good estimate of the noise with eight modes and has an
additional advantage in computing speed.  Fitting a narrow frequency
range of $\sim 40$ MHz around the water line with a high-order
polynomial tends to underestimate the noise level determined from the
whole band by about a factor of two.  However, the difference introduced
by the baseline fitting method is smaller than the uncertainty in the
upper limit for the outgassing rate due to unknown model parameters, and
therefore our data reduction method does not modify the conclusions.
The EMD and HHT reduction methods applied to the baseline removal and
denoising of the {\it Herschel}/HIFI data will be described in detail
in a future work (Rezac et al.\ in preparation).  We show the baseline
subtracted HRS spectrum in Fig.~\ref{fig:model} with the expected line
emission overplotted.

\section{Outgassing rate}

There is no evidence of \ce{H2O} emission in our observation, although
it is expected that the object's dust emission activity is driven by the
sublimation of subsurface material as it approaches perihelion.  A
molecular excitation model based on the publicly available accelerated
Monte Carlo radiative transfer code {\it ratran}
\citep{2000A&A...362..697H} is used to calculate the population of the
rotational levels of water as a function of the nucleocentric distance.
The code includes collisional effects and infrared fluorescence by solar
radiation to derive the production rates. We used the one-dimensional
spherically symmetric version of the code following the description
outlined in \citet{2004ApJ...615..531B} that has been used to analyze
\herschel{} and ground-based cometary observations \citep[see
e.g.][]{2010A&A...518L.150H,2010A&A...521L..50D,2011Natur.478..218H,2012arXiv1208.4358D}.
The model input parameters are the gas kinetic temperature, which
controls the molecular excitation in the collisional region, and the
electron density.  We assume a gas kinetic temperature in the range
20--40 K.  The electron density and temperature profiles from
\citet{1997PhDT........51B} are adopted.  Since the electron density in
the coma is not well constrained, an electron density scaling factor of
$\xne = 0.2$ with respect to the standard profile derived from
observations of comet 1P/Halley has been used
\citep[e.g.][]{2010A&A...518L.150H}.  The expansion velocity is assumed
to be constant in the coma, and the radial gas density profile for water
was obtained using the standard spherically symmetric Haser
distribution.  The gas expansion velocity derived from the pre- and
post-perihelion evolution of comet C/1995 O1 (Hale–Bopp) is given by
$v_\mathrm{exp} = 1.12 \times \rh^{-0.41}$ \kms
\citep{1997EM&P...78....5B}, which corresponds to 0.76 \kms at 176P's
heliocentric distance. Since Hale-Bopp was a very active comet, this
expansion velocity is a very conservative upper limit.  A thermal
velocity of 0.35 \kms{} is obtained from the temperature expected at the
subsolar point where ice sublimates. Then, as cometary atmospheres are
formed by quasi-adiabatic expansion, the gas accelerates as it expands,
as observed with the large field of view of the 18-cm OH observations.
For low-activity and distant comets an expansion velocity close to 0.5
\kms{} is determined from the shapes of the OH line observed with the
Nançay radio telescope \citep{2007A&A...467..729T}, but observations for
$Q_\ce{OH} < 10^{28}\ \s$ are lacking, and observations at $\rh > 2$ AU
are rare.  Odin observations of the \ce{H2O} 557 GHz line toward the
active comet C/2003 K4 (LINEAR) at 2.2 AU from the Sun are consistent
with an expansion velocity on the order of 0.5 \kms
\citep{2007P&SS...55.1058B}.  The dust particles ejected by 176P have an
approximate velocity or $\sim5$ \ms calculated from numerical
simulations to match optical observations \citep{2011AJ....142...29H}.
However, dust is expected to be much slower than gas if it is formed by
large and heavy particles with very low outgassing rates.

The mean of the derived upper limits on the total \ce{H2O} column
density integrated within the beam for the considered model parameters
is $\langle N_\ce{H2O} \rangle < \column \times10^{10}\
\mathrm{cm}^{-2}$.  For a low expansion velocity characteristic of weak
comets, we derive a sensitive 3-$\sigma$ upper limit on the water
production rate of $< 2.1 \times 10^{25}\ \s$ from the WBS and HRS data
(see Table~\ref{tbl:q}).  An upper limit of $< \upper \times 10^{25}\
\s$ is derived from the mean of the WBS and HRS upper limits for gas
expansion velocities between 0.4--0.7 \kms and gas kinetic temperatures
between 20--40 K.

With the exception of 133P/Elst-Pizarro, our upper limit for the
\ce{H2O} production rate is more stringent than any of those derived in
other MBCs from the spectroscopic search for CN emission in the optical
when the objects were active.  In addition, it does not require the
uncertain assumption of a $Q_\ce{CN}/Q_\ce{H2O}$ value.  Adopting a
$Q_\ce{CN}/Q_\ce{H2O}$ mixing ratio of $10^{-3}$ the upper limits are
$Q_\ce{H2O} < 1.3 \times 10^{24}\ \s$ in 133P/Elst-Pizarro
\citep{2011A&A...532A..65L}, $Q_\ce{H2O} < 1.4 \times 10^{26}\ \s$ in
P/2008 R1 (Garradd) \citep{2009AJ....137.4313J}, $Q_\ce{H2O} < 1.3
\times 10^{27}\ \s$ in P/2006 VW139 \citep{2012ApJ...748L..15H},
$Q_\ce{H2O} < 6 \times 10^{26}\ \s$ in P/2010 R2 (La Sagra)
\citep{2012AJ....143..104H}, and $Q_\ce{H2O} < 9 \times 10^{26}\ \s$ in
the collisionally disrupted main belt object (596) Scheila
\citep{2012ApJ...744....9H}. However, no direct search for \ce{H2O} has
been carried out before, and there are substantial uncertainties in the
estimation of the water production from the CN emission given the wide
range of observed $Q_\ce{CN}/Q_\ce{H2O}$ ratios in comets and their
dependence on heliocentric distance.  Therefore, these previously
published values can only be considered as order-of-magnitude
approximations.

\section{Discussion}

\begin{table}
  \caption{Standard deviation of the brightness temperature and line
  area, and retrieved 3-$\sigma$ upper limits of the \ce{H2O} production
  rate in 176P/LINEAR.}
  \label{tbl:q}
  \centering
  \begin{tabular}{c c c c}
    \hline\hline
    Spectrometer & $\sigma_{T_\mathrm{mB}}$ & $\sigma_{\int T_\mathrm{mB}\, dv}$ & $Q_\ce{H2O}$\tablefootmark{a} \\
    & (K) & (K \kms) & (\s) \\
    \hline
    WBS & $6.598 \times 10^{-4}$ & $6.172 \times 10^{-4}$ & $< 2.08 \times 10^{25}$ \\
    HRS & $1.998 \times 10^{-3}$ & $6.365 \times 10^{-4}$ & $< 2.14 \times 10^{25}$ \\
    \hline
  \end{tabular}
  \tablefoot{
  \tablefoottext{a}{Production rates derived for a gas kinetic
  temperature of 20 K, expansion velocity of 0.5 \kms, and an electron
  density scaling factor of $\xne = 0.2$.}
  }
\end{table}

We observed water emission in 176P with
\herschel/HIFI to test the prediction that cometary activity in MBCs is
driven by sublimation of water ices from the nucleus.  There are several
mechanisms that have been proposed to drive mass loss from small bodies,
including sublimation of subsurface ices, rotational instability,
impact ejection and thermal fracture \citep[see][for recent reviews of
MBC physical properties and activation
mechanisms]{2011P&SS...59..365B,2012AJ....143...66J}.  The cometary
activity observed in 176P was initially found to suggest the
presence of sublimating subsurface ice that may have been exposed by
recent collisions \citep{2011AJ....142...29H}.  This view is supported
by the detection of water ice absorption in spectroscopic observations
centered at 3.1 $\mu$m of the surface of the largest asteroid of the
Themis asteroid family, 24 Themis, which belongs to the same dynamical
family as 176P \citep{2010Natur.464.1322R,2010Natur.464.1320C}, although
it has been claimed that the measured spectra are consistent with the
transmission spectra of goethite \citep{2011A&A...526A..85B}.  From the
search for the \ce{H2O} \trans{} rotational line at 557 GHz in 176P, a
3-$\sigma$ upper limit for the \ce{H2O} production rate of $< \upper
\times 10^{25}\ \s$ is derived from the WBS and HRS data, for gas
expansion velocities between 0.4--0.7 \kms and gas kinetic temperatures
between 20 and 40 K.  Using the peak value of the $R$-band magnitude, $m(1,
1, 0) = 15.35 \pm 0.05$, measured when the object was active in late
2005 at a heliocentric distance of 2.58 AU, a $V$-band magnitude of
$m_V(1, \rh, 0) = 17.8$ is obtained \citep{2009ApJ...694L.111H}.  Since
the cometary activity in 176P is indicative of ice sublimation with a
30\% contribution of the coma to the total brightness, the scaling
relation between gas production rates and heliocentric magnitudes from
\citet{2008LPICo1405.8046J} predicts a water production rate of
approximately $1.0 \times 10^{26}\ \s$.  This correlation has been
obtained for a sample of 37 comets with heliocentric distances between
0.32--4.53 AU.  Thus, if ice sublimation is the driving mechanism of
176P's activity, the derived \ce{H2O} production rate is too low by
about a factor of two to explain the activity level during its 2005
perihelion passage.  It is unlikely that sublimation of carbon monoxide
ice is the source of the activity in MBCs since the temperature in the
region where the comet formed could not have allowed condensation of CO.
We note that the \citet{2008LPICo1405.8046J} correlation should be taken
with some care because none of the comet measurements were obtained at
the brightness level of 176P.  From the dust production rate of 0.07 kg
s$^{-1}$ estimated by \citet{2011AJ....142...29H}, a water production
rate of $2.3 \times 10^{24}\ \s$ is inferred assuming a dust-to-gas
ratio of one.  However, there are significant uncertainties in the value
of the dust production rate determined from photometric data, and
additionally the dependence on the dust-to-gas ratio with heliocentric
distance is poorly constrained.  Mid-infrared photometric observations
of 176P on 23--24 April 2010 by the {\it Wide-field Infrared Survey
Explorer} (WISE) mission did not provide any indication of a coma
\citep{2012ApJ...747...49B}, and there are no other published infrared or
optical data closer to the last perihelion passage.  We conclude that
water was not detected in our observation because the water production
rate was lower than $\upper \times 10^{25}\ \s$ or the object was not
active during our observation, so a more detailed study is needed to
shed light on the activation mechanism in MBCs.

\begin{acknowledgements}
HIFI has been designed and built by a consortium of institutes and
university departments from across Europe, Canada, and the United States
under the leadership of SRON, Netherlands Institute for Space Research,
Groningen, The Netherlands, and with major contributions from Germany,
France, and the US.
HIPE is a joint development by the \herschel{} Science Ground Segment
Consortium, consisting of ESA, the NASA \herschel{} Science Center, and
the HIFI, PACS, and SPIRE consortia.
Support for this work was provided by NASA through an award issued by
JPL/Caltech. MdVB was supported by the Special Priority Program 1488 of
the German Science Foundation.  SS was supported by Polish MNiSW funds
(181/N-HSO/2008/0). We acknowledge the referee, H. Campins, for his
comments.
\end{acknowledgements}

\bibliographystyle{aa}
\bibliography{ads,ladi,preprints}

\end{document}